\documentclass[twocolumn,showpacs,preprintnumbers,amsmath,amssymb]{revtex4}
\usepackage{dcolumn}
\usepackage{bm}
\usepackage{graphicx}
\usepackage{amsmath}
\usepackage{subfigure}
\usepackage{changes}
\usepackage{graphicx}
\usepackage{amsmath}

\begin{document}

\title{Robust external spin hyperpolarization of quadrupolar nuclei enabled by strain}
\author{Lu Chen$^1$, Jiawen Jiang$^1$, Martin B. Plenio$^2$ and Qiong Chen$^1$\footnote{E-mail:qchen@hunnu.edu.cn}}
\affiliation{$^1$Key Laboratory of Low-Dimension Quantum Structures and Quantum Control of Ministry of Education, \\
Synergetic Innovation Center for Quantum Effects and Applications, Xiangjiang-Laboratory and Department of Physics, \\
Hunan Normal University, Changsha 410081, China\\
$^2$Institut f\"{u}r Theoretische Physik and IQST, Albert-Einstein-Allee 11, Universit\"{a}t Ulm, D-89081 Ulm, Germany}
\begin{abstract}
In a theoretical study, we investigate the spin dynamics of interacting nitrogen-vacancy (NV) centers and quadrupolar I = 3/2 nuclear spins, specifically ${}^{11}$B spins in hexagonal boron nitride (h-BN) nanosheets located near the microdiamond surface. We demonstrate the possibility of obtaining external spin-polarization by magnetic field sweeps across the level anticrossings around zero field. To achieve this, we leverage crystal strains to establish a polarization transfer mechanism that remains robust against variations in NV orientation, crystal strain inhomogeneity, and electron-nuclear effective couplings. These results pave the way for hyperpolarization of spins in nanomaterials near the diamond surface without experiencing polarization loss to intrinsic nuclear spin-1/2 species, such as ${}^{13}$C and ${}^{1}$H nuclear spins in diamond. The ${}^{11}$B spins in h-BN nanosheets, with their extended relaxation time and large surface area, present a promising alternative for relayed nuclear polarization to the liquid phase and for the development of quantum simulators based on surface nuclear spins.
\end{abstract}
\maketitle

\emph{Introduction ---}
Nuclear magnetic resonance (NMR) is a powerful versatile technique to address a wide range of fields in chemistry, biology, and medicine. Conventional NMR detection performed with inductive coils, requires high magnetic field and large detection volume to improve the signal-to-noise ratio (SNR) and spectral resolution. Nuclear polarization is determined by thermal (Boltzman) polarization, which is exceedingly low, around 10$^{-5}$ for hydrogen spins and 3 T magnetic field at room temperature. Hence, several hyperpolarization methods have been proposed to enhance this polarization \cite{BowersW1986,BowersW1987,ArdenkjaerFG+2003,eichhorn2022hyperpolarized}.  Nitrogen-vacancy (NV) centers in diamond \cite{doherty2013nitrogen,jelezko2006single,wu2016diamond}, are examples of optical hyperpolarizing agents that can achieve almost perfect spin polarization through optical pumping, independent of the magnetic field, enabling hyperpolarization on nearby nuclei in bulk diamond \cite{london2013detecting,fischer2013bulk,jacques2009dynamic,alvarez2015local}, micro and nanodiamonds \cite{miyanishi2021room,ajoy2020room,rej2017hyperpolarized}.

The investigation of optically “hyperpolarized nanodiamonds” is a long-standing and highly interesting research area because nanoscale or microscale particles of diamond in powdered form offer a significantly larger contact surface area \cite{boele2020tailored}. Several control protocols have been proposed for ${}^{13}$C spin hyperpolarization intrinsic to the micro and nanodiamond, such as PulsePol sequences \cite{schwartz2018robust}, microwave (MW) \cite{chen2015optical,ajoy2018orientation,ajoy2018enhanced,zangara2019dynamics,ajoy2021low} or magnetic field sweeps \cite{henshaw2019carbon,zangara2019two,wunderlich2021robust}. These approaches have been demonstrated in recent proof-of-principle experiments. However, achieving bulk polarization is only the first step. Subsequent stages involve the slow process of polarization diffusion toward the surface and then the transfer to external spins in the liquid phase. Unfortunately, these steps can be hindered by the presence of P1 centers and other paramagnetic impurities in diamond. In other words, the transfer of $^{13}$C polarization to external spins is prevented by dark spins on the surface.

On the other hand, while some evidence for the efficient transfer of polarization from an NV center, which lacks strong couplings, to a dense layer of nuclear spins has already been provided with single crystalline diamond \cite{broadway2018quantum,healey2021polarization,2018toward}, no external spin hyperpolarization has been demonstrated yet, mainly due to low diamond surface-to-volume ratios. Although there have been theoretical advancements in polarization transfer to external nuclear spins in contact with powdered diamond \cite{chen2016resonance}, several significant challenges have been identified. These challenges include: large broadening of the electronic linewidth due to unavoidable misalignment between the applied magnetic field and the NV symmetry axis, relatively large distance between NV centers and target nuclei, short decoherence time of NV spins in small sizes of nanodiamonds  \cite{tisler2009fluorescence} or the lack of nuclear spin species selectivity of polarization protocols. In this study, we propose a protocol to overcome these challenges and achieve robust polarization buildup in quadrupolar nuclei in nanomaterials near the microdiamond surface. Specifically,  we focus on $I=3/2$ $^{11}B$ boron spins in hexagonal boron nitride (h-BN) \cite{love2018probing,lovchinsky2017magnetic,henshaw2022nanoscale} nanosheets, utilizing a shallow NV center ensemble (2$\sim$5 nm depth to diamond surface) at zero-to-low-field.

\begin{figure*}
\center\includegraphics[width=6.6in]{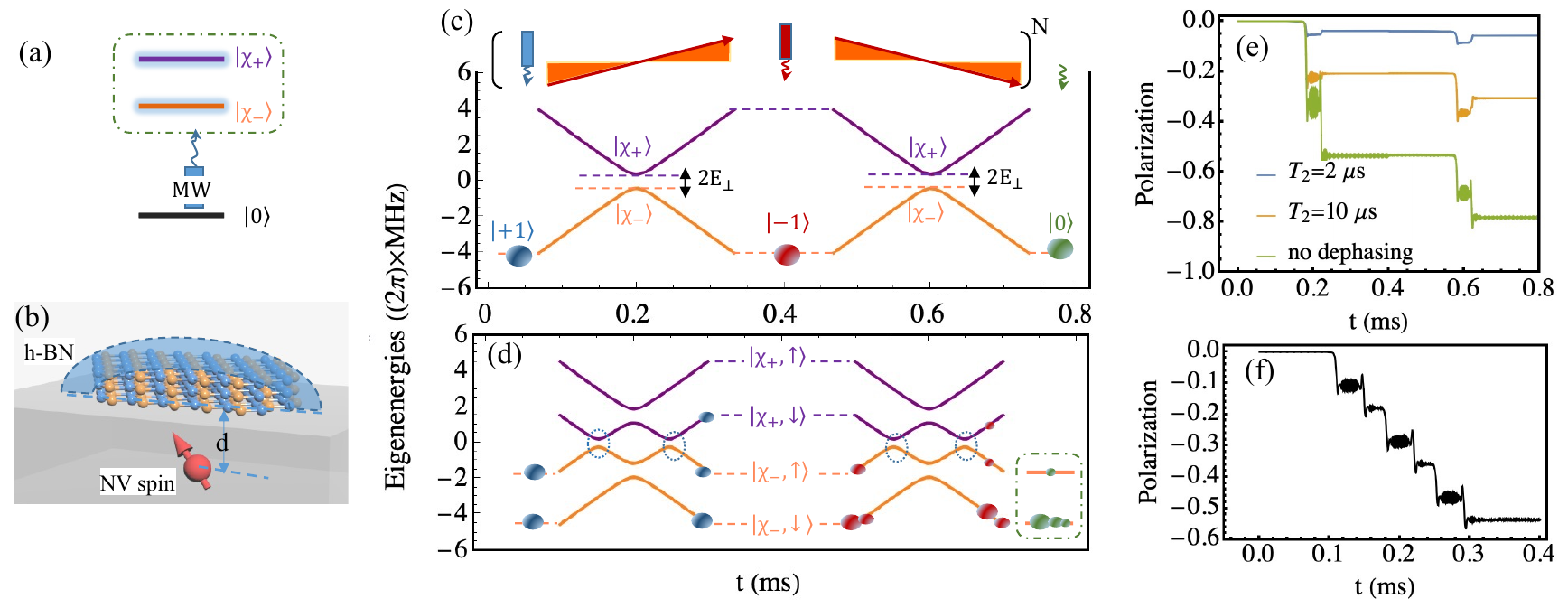}
\caption{Basic protocol of $^{11}B$ spin polarization mechanism around zero-field.  (a) The initial polarization subspace of the NV spin. (b) A shallow NV spin, at depth $d$ below the diamond surface, is interacting with a h-BN nanosheet near the diamond surface. (c) Energy diagram of an isolated NV electron spin for a full magnetic field cycle based on Eq. (1), which includes two magnetic field sweeps, from negative to positive, and then back to negative field strength (orange). Squares in different colors indicate laser pulses and a microwave (MW) pulse to initial NV spin in state $|\chi_{-}\rangle$ ($|\chi_{-}\rangle\sim|m_s=+1\rangle$ (blue) or $|\chi_{-}\rangle\sim|m_s=-1\rangle$ (red)). (d) Schematic energy diagram for the NV and a $^{11}B$ nuclear spin based on Eq. (2). Spin populations are represented by solid ellipses of variable radius; for clarity, we assume the NV spin is always fully initialized into $|\chi_{-}\rangle$, though only partial initialization is required. In both of the sweeps, there are two avoided crossings where population transfer takes place between $|\chi_{-},\uparrow\rangle$ and $|\chi_{+},\downarrow\rangle$. (e) Polarization of $^{11}B$ nuclear spins within a sweep cycle is shown by considering different coherence time $T_{2}$ of the NV spin and the duration of the LZ transition is $\tau\approx2$ $\mu$s. The quadrupole splitting constant $\bar{Q}=(2\pi)\times2.9$ MHz, the strength of crystal strain $E_{\perp}=(2\pi)\times0.4$ MHz, the effective perpendicular coupling between NV and nuclear spins $a_{x}=(2\pi)\times47$ kHz (the NV depth is assumed as 2 nm), the sweep rate $v=(2\pi)\times30$ MHz/ms and $\omega(t=0)=-(2\pi)\times6$ MHz. (f) The effect of $^{14}$N nuclear spins is included. The other parameters are the same during simulations for different color lines in (e). }
\label{protocol}
\end{figure*}

Unlike traditional methods that ignore crystal strain, in this study we employ crystal strain near diamond surfaces as an asset that imparts robustness. The crystal strain, is induced by local deformation of the diamond crystal, resulting in inhomogeneous broadening, typically in the range of 100 kHz in single-crystal diamond. However, in diamond powder and in shallow NV defects hosted in nanodiamonds, the inhomogeneous broadening can be even larger, reaching a few MHz \cite{Trusheim,marshall2022high,sharma2018imaging}. Remarkably, our protocol avoids polarization loss to other nuclear spin-1/2 species and achieves robust polarization transfer to $^{11}$B spins over a broad set of strain inhomogeneity, electron-nuclear couplings, and relative orientations of the NV axis and magnetic fields. The exceptional long relaxation time of $^{11}B$ spins, that can reach $T_{1b}=175$ s \cite{love2018probing} in h-BN nanosheets with thickness $\sim$50 - 100 nm and diameter 1 $\mu$m as well as the large surface area of the nanosheets benefit our polarization scheme. They may also provide a mediator to transfer the polarization to external molecules in solution. Additionally, our work provides a method for the initialization of quantum simulators based on surface nuclear spins \cite{tabesh2023active,cai2013large,gao2022nuclear}.





\emph{The model ---} Our target samples are $^{11} $B spins ($I=3/2$) in hexagonal boron nitride (h-BN) nanosheets at the diamond surface interacting with shallow NV spins ($\sim2 - 5$ nm), see Fig. 1(b). As a reasonable basic model we use an NV center interacting with a nearby $^{11}$B nuclear spin with the effective coupling $a_{x}\in(2\pi)\times$[0.01,0.05] MHz \cite{tetienne2021prospects,SI}. For clarity, we initially assume that the magnetic field is aligned with the NV orientation, and the applied magnetic field is sufficiently small to induce degenerate states $|m_I=\pm1/2\rangle$ and $|m_I=\pm3/2\rangle$ of the $^{11}$B nuclei.

We initialize the NV center in a highly polarized state in the $|m_s=\pm 1\rangle$ manifold, and gradually sweep the magnetic field across zero to the opposite direction which will transfer the NV polarization to the pseudo two-level system $^{11}$B spins consisting of the of degenerate states $|\downarrow\rangle=|m_I=\pm1/2\rangle$ and $|\uparrow\rangle=|m_I=\pm3/2\rangle$. Close to zero magnetic field, we have
\begin{equation}
H_{NV}=\omega(t)\sigma_{z}+E_{\perp}\sigma_{x},
\end{equation}
where $\omega(t)=\omega_0+vt$ is determined by the magnetic field with the initial frequency $\omega_0$ and field sweep rate $v$, $E_{\perp}$ is the perpendicular component of the crystal strain, with $\sigma_{z}=|m_s=+1\rangle\langle m_s=+1|-|m_s=-1\rangle\langle m_s=-1|$ and $\sigma_{x}=|m_s=+1\rangle\langle m_s=-1|+|m_s=-1\rangle\langle m_s=+1|$. Eigenstates of the NV spin are given by $|\chi_{\pm}(t)\rangle=\pm\cos\frac{\eta}{2}|m_s=\pm1\rangle+\sin\frac{\eta}{2}|m_s=\mp1\rangle$ with $\sin\eta=E_{\perp}/\sqrt{\omega^{2}(t)+E_{\perp}^{2}}$.

Our scheme starts with a magnetic field initialized to an orientation opposite to that of the NV such that $\omega(t=0)=-(2\pi)\times6$ MHz. The NV spin is initialized into state $|m_s=0\rangle$ by using laser illumination and then transferred to eigenstate $|\chi_{-}\rangle$ approximated as $|m_s=+1\rangle$ through a suitably tuned microwave (MW) pulse as shown in Fig. 1(c). The same MW pulse will also induce state transfer to $|\chi_{-}\rangle$ approximated as $|m_s=-1\rangle$ when the initial magnetic field direction is along NV axis. The NV spin is reinitialized in state $|\chi_{-}\rangle$ after every sweep and a repetition of these processes gives negative polarization of $^{11}$B spins in state $|\downarrow\rangle$ thanks to a dynamics that will be described in the following.

According Landau-Zener (LZ) theory, when we initialize an isolated NV spin into state $|m_s=+1\rangle$ at the start of the sweep with $\omega(t=0)=-(2\pi)\times6$ MHz, a fully adiabatic slow passage, satisfying the adiabatic condition $4 E_{\perp}^2/v\gg 1$, would transfer the NV spin to state $|m_s=-1\rangle$ at the end of the sweep (see Fig 1(c)).
%
\begin{figure}
\center\includegraphics[width=3.2in]{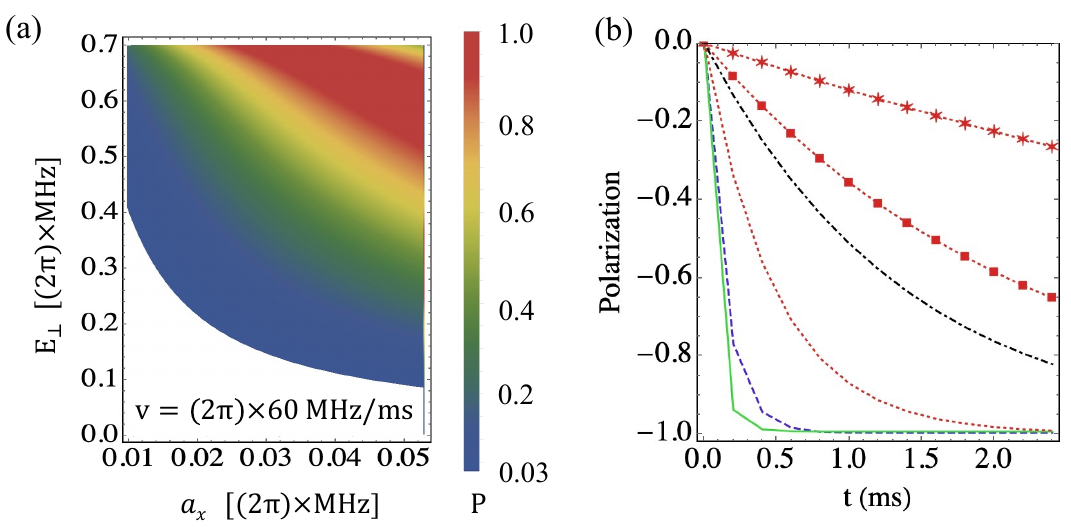}
\caption{The robustness of our scheme. (a) Transfer probability P as a function of the coupling strengths and crystal strains to show the robustness. (b) Polarization is transferred to a $^{11}$B spin by assuming $E_{\perp}=(2\pi)\times$0.65 MHz with $a_{x}=(2\pi)\times$ 0.012(black dot-dashed); 0.035(blue dashed); 0.047(green) MHz. The red dotted lines present $a_{x}=(2\pi)\times$ 0.02 MHz with $E_{\perp}=(2\pi)\times$0.65(no marks); 0.45(cubes); 0.25(stars) MHz.}
\label{protoco2}
\end{figure}

To understand the polarization mechanism, due to the coupling of the NV to the $^{11}$B nuclei during quasi-adiabatic sweep we consider the energy diagram of the NV-$^{11}$B system by using the NV-$^{11}$B coupling Hamiltonian as
\begin{align}
H_{1}=\omega_{e}\sigma_{\tilde{z}}+\frac{\bar{Q}}{2}I_{\tilde{z}}+H_{int},
 \end{align}
in which $\sigma_{\tilde{z}}=\frac{1}{2}(|\chi_{+}\rangle\langle\chi_{+}|-|\chi_{-}\rangle\langle\chi_{-}|)$ and $I_{\tilde{z}}=\frac{1}{2}(|\uparrow\rangle\langle\uparrow|-|\downarrow\rangle\langle\downarrow|$), $\omega_{e}=2\sqrt{\omega^{2}(t)+E_{\perp}^{2}}$, quadrupole coupling constant $\bar{Q}$=$(2\pi)\times2.9$ MHz and $H_{int}=2(\sigma_{\tilde{z}}\cos\eta-\sigma_{\tilde{x}}\sin\eta)\cdot (\sqrt{3}a_{x}I_{\tilde{x}}+2a_{z}I_{\tilde{z}})$ describes the dipole-dipole interaction between NV and  $^{11}$B spins. A full polarization transfer cycle includes the sequential application of negative-to-positive and positive-to-negative magnetic field sweeps, in which the NV spin is initialized into state $|\chi_{-}\rangle$ at each sweep start.

By initializing the NV spin into state $|\chi_{-}\rangle$, for each magnetic field sweep, the state $|\chi_{-},\downarrow\rangle$ is not affected as its dynamics remains adiabatic ($4 E_{\perp}^2/v\gg 1$), while the initial state  $|\chi_{-},\uparrow\rangle$  will have a probability for flipping to $|\chi_{+},\downarrow\rangle$. This is the crucial key of the polarization transfers, when Hartmann-Hahn condition
\begin{equation}
\omega_{e}=2\sqrt{\omega^{2}(t)+E_{\perp}^{2}}= \frac{\bar{Q}}{2}
\end{equation}
could be matched at each of avoided crossings. As shown in Fig. 1(d), if we have $0<|E_{\perp}|<\frac{\bar{Q}}{4}$, there are sequential two avoided crossings of the energy levels for a negative-to-positive or positive-to-negative sweep, and the energy gap of the non-zero-field avoided crossing is given by $\sqrt{3}a_{x}\sin\eta$. The sweep rate is chosen to be partially non-adiabatic Landau-Zener (LZ) dynamic around the avoided crossing (blue dotted ellipses in Fig.1(d)) with LZ transition probability $p_l\sim\exp(-2\pi \mu)$ with $\mu=3a_{x}^2\sin^{2}\eta/(8v\cos\eta)$ between the eigenstates $|\chi_{-},\uparrow\rangle$ and $|\chi_{+},\downarrow\rangle$.This probability is given by $P_l=P\sin^2[\Phi]$ with
\begin{equation}
P=4p_l(1-p_l),
\label{P}
\end{equation}
and the phase $\Phi$ is related to parameters in both of adiabatic and non-adiabatic evolutions in the system \cite{SI,ashhab2007two,shevchenko2010landau}. Therefore the net nuclear polarization build-up depends on the coherent population transfer around every avoided crossings due to partially non-adiabatic LZ transition probabilities. For an initial state of the NV given by $|\chi_{-}\rangle$ and assuming the $^{11}$B nuclei unpolarized, the polarization cycle results in negative polarization of $^{11}$B spins in state $|\downarrow\rangle$ regardless of the magnetic field sweep direction, as confirmed by direct numerical simulations both for negative-to-positive and positive-to-negative field sweeps (as shown in Fig. 1(e)). Initialization of the NV spin into state $|\chi_{+}\rangle$ induces positive nuclear polarization to state $|\uparrow\rangle$.

\emph{Robustness ---} Regarding each avoided crossing as a coherent process, as shown in Fig. 1(e), polarization is transferred under the condition $\tau< T_2(NV) $, where $T_2(NV)$ denotes the NV coherence lifetime, and $\tau\sim\sqrt{3}a_x|\sin\eta|/(2v|\cos\eta|)$ is the characteristic LZ time to coherent transfer. An accurate description of the system spin dynamics during a single magnetic field sweep, in which the coupling of NV and $^{14}$N nuclear spin is included, is presented in Fig. 1(f). We find that the interactions between NV and $^{14}$N nuclear spins do not change the efficiency of nuclear polarization mechanism of $^{11}$B nuclei (detailed calculations are included in supplemental materials \cite{SI}).


In practice, polarization transfer will involve a large ensembles of NV spins and target $^{11}B$ nuclei. We study this situation by considering the case of polarization of nuclear spins in h-BN nanosheets on a microdiamond surface with an ensemble of shallow NV centers. Because the Stoke phase $\Phi$ is sensitive to system parameters and is assumed to be averaged as $\langle\sin^2[\Phi]\rangle\approx1/2$ \cite{shevchenko2010landau,chen2015optical}, we calculate $P$ as a function of both the strain components and the NV-$^{11}$B couplings to estimate the robustness of the parameter ranges, as shown in Fig. 2(a) and supplemental material \cite{SI}. For simplicity, we first assume the magnetic field is along NV's orientation. Remarkably, according to our theory,  it shows polarization transfer when $E_{\perp}$ is comparable to but not larger than $\frac{\bar{Q}}{4}$. Therefore, we have polarization transfer over a broad set of NV-$^{11}$B effective couplings and strain components, which is confirmed by our exact numerical simulations, as shown in Fig. 2(b).

Finally, we investigate the nuclear polarization efficiency as a function of the magnetic field orientation relative to the NV centers in diamond powders, in which the Hamiltonian of an NV spin is
\begin{equation}
H_{\theta}=[\delta_b+\omega_{\theta}(t)]\sigma_{z}+E_{\perp}\sigma_{x},
\end{equation}
here $\omega_{\theta}(t)=\omega(t)\cos\theta$ with angle $\theta$ between NV's orientation and magnetic field, and $\delta_b$ is  inhomogeneous broadening induced by interactions with electron and nuclear spins nearby the NV center. We assume $v=(2\pi)\times60$ MHz/ms and average 300 runs to have the averaged polarization transfer of $^{11}$B spins by considering different couplings to the NV spin, inhomogeneities of the longitudinal and transversal component of the crystal strain $E_{z}\in(2\pi)\times[-0.73, 0.73]$ MHz, $E_{\perp}\in(2\pi)\times[-1.5, 1.5]$ MHz and $\delta_b\in(2\pi)\times[-3, 3]$ MHz of approximate Gaussian probability distributions $p(g)=\frac{1}{\sqrt{2\pi}\sigma_{g}}\exp(-g^2/2\sigma_{g}^2)$ with $\sigma_{g}/(2\pi)=0.25$, $0.5$ or $1$ MHz corresponding to $g=E_{z}$, $E_{\perp}$ or $\delta_b$. As shown in Fig. 3(c), the spin dynamics is insensitive to the exact start and end magnetic field values, numerical simulations of the averaged polarization transfers show similar performances regardless of NV orientation variance and $\delta_b$ values \cite{SI}. The MW initialization of polarization NV centers, defined as population difference between states $|+1\rangle$ and $|-1\rangle$ \cite{SI}, is also a limiting factor of the final polarization, as shown in Fig. 3(b), although inhomogeneous broadening $\delta_b$ and longitudinal component of the crystal strain $E_{z}$ has almost no effect in the polarization transfer in our method,  they are related to polarization initialization of the NV centers. We have $P_{NV}>0.1$ when $\theta=80^{\circ}$, and therefore conclude our method is confined to the solid cone defined by $\theta>80^{\circ}$, which includes $83\%$ of NV orientations.

By assuming a solid mixture of microdiamonds and h-BN nanosheets with the same volume concentrations as 1:1, we can estimate the final polarization $\bar{P_b}\approx\rho_{NV}T_{o}P_1/\rho_n$ with $T_o=100$ s the total cycle time, NV density $\rho_{NV}=1.6\times10^{4}$ $\mu$m$^{-3}$ and $^{11}$B spin density $\rho_{n}=1.6\times10^{10}$ $\mu$m$^{-3}$. When $d\sim2$ nm, our mechanism gives polarization rate as $P_1\sim0.17/(0.2ms)$ (see Fig. 3(b)), and we have the final polarization $\bar{P_b}\approx8.5$\%. Similar estimation of the $d\sim5$ nm case gives 2\% of $^{11}$B spins to be polarized. Although we study of $^{11}$B spins in h-BN nanosheets, our method could be used for polarization of the other quadrupolar nuclei and materials, i.e., $^{27}$Al spins in Al$_2$O$_3$ nanoparticles.


 %
\begin{figure}
\center\includegraphics[width=3.2in]{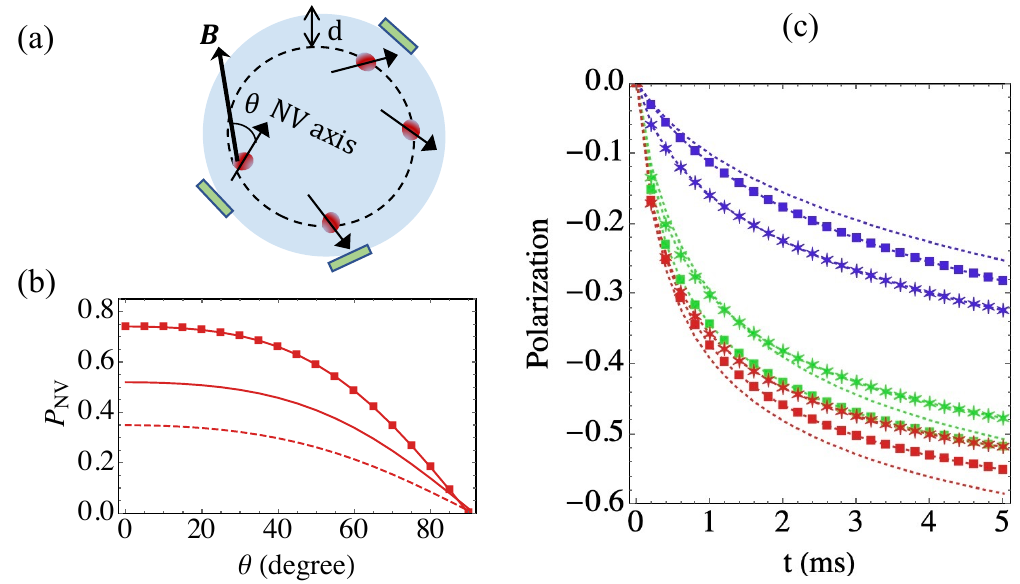}
\caption{The polarization build-up of $^{11}$B spins in h-BN nanosheets by considering a solid mixture of microdiamonds (diameter $>1 \mu$m) and h-BN nanosheets  (diameter $<1 \mu$m). (a) NV centers are $d\sim$2-5 nm depth from the microdiamond surface and h-BN nanosheets are covered on the surface. (b) The ensemble-averaged initial polarization of the NV centers regarding the angle $\theta$ between their orientations and magnetic field by assuming 300 times averaging in random values within $E_{z}\in(2\pi)\times[-0.73, 0.73]$ MHz and and $\delta_b=(2\pi)\times$0(red cubes), $\in(2\pi)\times$[-2, 2](red solid), [-3, 3](red dashed) MHz, MW Rabi frequency is given by $(2\pi)\times8$ MHz and the
ensemble-averaged initial polarization is an average over all the possible angles between NV axis and MW field \cite{SI}. (c) Polarization transfer simulation by assuming 300 times averaging in random values within $E_{\perp}\in(2\pi)\times[-1.5, 1.5]$ MHz and $\delta_b\in(2\pi)\times[-3, 3]$ MHz. Three different angles are given $\theta=0^{\circ}$ (dashed line), $40^{\circ}$ (cubes) and $75^{\circ}$ (stars) corresponding to couplings $a_{x}=(2\pi)\times$ 0.012(blue); 0.026(green); 0.047(red) MHz.}
\label{protoco2}
\end{figure}
Before concluding, we would like to mention that our polarization mechanism shares some similarities with other methods of MW sweeps and field sweeps. However, in our scheme, NV spin works as a pseudo two-level system constituted of states $|m_s=\pm1\rangle$ so that there is no similar hierarchical gaps of NV and $^{13}$C spins encountered in Ref. \cite{ajoy2018orientation}. Moreover, we take advantage of the transverse zero-field splitting induced by the strain to impart robustness.  The principles of our protocol allow for extension to other systems where transverse zero-field splitting exists and thus open a much wider field of application. Specifically, molecular color centers, has gained growing interest in recent years \cite{bayliss2020optically,wasielewski2020exploiting}, owns a noticeable transverse zero-field splitting due to low symmetry. We provide a general method to polarize quadrupolar nuclei in nanomaterials out of the diamond without the polarization loss to $^{13}$C spins in diamond, which opens another way for chemical and structural analysis of nanomaterials or other compounds surface layers on microdiamonds through enhancing NMR spectroscopy around zero field. Low magnetic field gives a relative small sweep range, that leads to fast polarization transfers, and $83\%$ of NV orientations giving contributions to polarization transfer.  Although low magnetic field leads to a smaller relaxation time of nuclei, comparing with previous sweep schemes to polarize $^{13}$C spins in microdiamonds, paramagnetic P1 centers in diamond reduce relaxation time $T_{1c}$ of $^{13}$C spins in diamond dramatically \cite{T1C13} but have small effects to $^{11}B$ spins due to their large quadrupolar couplings.


 \emph{Conclusion ---} In summary, we present a nuclear polarization mechanism applicable to diamond powder mixtures, enabling direct polarization of quadrupolar nuclear spin targets outside the diamond crystal instead of $^{13}$C spins within diamond. Central to this approach is utilization of Landau-Zener dynamics induced by quasi-adiabatic magnetic field sweeps across the set of avoided crossings, which arise from nearly matched energy differences of the individual NV and  $^{11}$B spins. This technique can operate under ambient conditions, and remains robust to strain inhomogeneity, spin coupling heterogeneity and NV orientation disorder. Hyperpolarization of $^{11}$B spins in nanosheets offers a combination of long relaxation time and large surface area, making it an excellent alternative for a mediated nuclear polarization medium. Our work introduces another promising platform for exploiting NV based hyperpolarization to achieve optically hyperpolarized nanomaterials in solid phase and in a liquid at room temperature.

{\em Acknowledgements --}
We thank Fedor Jelezko for comments on the manuscript. Q. Chen are supported by Natural Science Foundation of China (Grants No. 12375012, 12247105), the Science $\&$ Technology Department of Hunan Province (2023ZJ1010), and Hunan provincial major sci-tech program  (XJ2302001). M.B. Plenio was supported by the European Research Council Synergy Grant HyperQ (grant no. 856432) and the BMBF Zukunftscluster QSense: Quantensenoren f{\"u}r die biomedizinische Diagnostik (QMED) (grant no 03ZU1110FF).

Lu Chen and Jiawen Jiang contributed equally to this work.

\end{document}


\title{Supplementary materials for ``Robust external spin hyperpolarization of quadrupolar nuclei enabled by strain"}
\author{Lu Chen$^1$, Jiawen Jiang$^1$, Martin B. Plenio$^2$ and Qiong Chen$^1$\footnote{E-mail:qchen@hunnu.edu.cn}}
\affiliation{$^1$Key Laboratory of Low-Dimension Quantum Structures and Quantum Control of Ministry of Education, \\
Synergetic Innovation Center for Quantum Effects and Applications, Xiangjiang-Laboratory and Department of Physics, \\
Hunan Normal University, Changsha 410081, China\\
$^2$Institut f\"{u}r Theoretische Physik and IQST, Albert-Einstein-Allee 11, Universit\"{a}t Ulm, D-89081 Ulm, Germany}
\maketitle
\section{The effective Hamiltonian}
The Hamiltonian of the whole system is given by
\begin{align}
H&=H_{NV}+H_{^{11}B}+H_{^{14}N}+H_{P1}+H^{'}_{^{14}N}+H_{dip(NV,^{11}B)}\\\nonumber&+H_{dip(NV,^{14}N)}+H_{dip(NV,P1)}+H_{dip(^{11}B,P1)}+H^{'}_{dip(NV,^{14}N)},
\end{align}
in which the $H_{NV}$, $H_{^{11}B}$, $H_{^{14}N}$, $H_{P1}$ and $H^{'}_{^{14}N}$ correspond to the Hamiltonian of an nitrogen-vacancy (NV) center, the target $^{11}B$ nuclear spins in Hexagonal boron nitride (h-BN) nanosheets, the host $^{14}N$ nuclear spin of the NV in diamond, the substitutional nitrogen (P1) centers, and the $^{14}N$ nuclear spins in h-BN nanosheets, respectively. The terms $H_{dip(NV,^{11}B)}$, $H_{dip(NV,^{14}N)}$, $H_{dip(NV,P1)}$, $H_{dip(^{11}B,P1)}$, and $H^{'}_{dip(NV,^{14}N)}$ correspond to the dipole-dipole interactions Hamiltonian between the NV center and the $^{11}B$ nuclear spins, the NV center and its host $^{14}N$ nuclear spin, the NV center and the P1 center electron spins, the $^{11}B$ nuclear spins and the P1 center electron spins, and the NV center and $^{14}N$ nuclear spins in h-BN nanosheets. We will demonstrate in Section \ref{Eq6} that the dipole-dipole interactions between the NV center and other spins (P1 centers, the host $^{14}N$ nuclear spin, the $^{14}N$ nuclear spins in h-BN nanosheets) do not affect the polarization transfer efficiency in our scheme, in order to illustrate the mechanism, here we consider an NV center and a target $^{11}B$ spin, and the Hamiltonian is written as
\begin{align}
H_{m}&\simeq \overbrace{(D+E_{z})S^{2}_{z}+E_{x}(S_{x}^{2}-S_{y}^{2})+E_{y}(S_{x}S_{y}+S_{y}S_{x})+\omega(t)S_{z}}^{H_{NV}}\\\nonumber&
+\underbrace{\frac{\bar{Q}}{4I(2I-1)}[3I_{z}^{2}-I^{2}+\zeta(I_{x}^{2}-I_{y}^{2})]}_{H_{^{11}B}}+\underbrace{\vec{S}\cdot\mathcal{A} \cdot\vec{I}}_{H_{dip(NV,^{11}B)}},
\end{align}
where $\mathcal{A}$ describes the dipole-dipole interaction tensor of the NV center with the $^{11}B$ nuclear spin, spin operator $\vec{S}=(S_{x},S_{y},S_{z})$ and $\vec{I}=(I_{x},I_{y},I_{z})$ correspond to the NV center electron spin ($S$=$1$) and the $^{11}B$ nuclear spin ($I$=$\frac{3}{2}$), $D$=$(2\pi)\times$2.87 GHz is the zero-field splitting parameter, and $E_{x,y,z}$ are the crystal strains in all directions. $\omega(t)$=$\omega_{0}+vt$ is determined by the magnetic
field $B$ and field sweep rate $v$. The quadrupole coupling constant is $\bar{Q}$=$(2\pi)\times$2.9 MHz and the asymmetry parameter $\zeta$=$0$ \cite{lovchinsky2017magnetic} for $^{11}B$ nuclear spins in h-BN nanosheets.

The magnetic dipole-dipole interaction between an NV center and a single $^{11}B$ nuclear spin $H_{dip(NV,^{11}B)}$ can be written as
\begin{align}
H_{dip(NV,^{11}B)}\simeq \frac{\mu_{0}\hbar\gamma_{e}\gamma_{n}}{4\pi R^{3}}S_{z}(A_{x}I_{x}+A_{y}I_{y}+A_{z}I_{z}),
\label{Eq5}
\end{align}
by assuming the secular approximation, where $\gamma_{e}$,$\gamma_{n}$ are gyromagnetic ratios of the NV center electronic spin and $^{11}B$ nuclear spin, respectively. The $R$ denotes the distance between the NV center and the nuclear spin, $\alpha$ and $\varphi$ define the angular orientation of $\vec{R}=(\vec{R}_{x},\vec{R}_{y},\vec{R}_{z})$. $A_{x}$=$3\sin\alpha\cos\alpha\cos\varphi$, $A_{y}$=$3\sin\alpha\cos\alpha\sin\varphi$ and $A_{z}$=$3\cos^{2}\alpha-1$. Then, we consider the effective transverse coupling $A_{x^{'}}$=$\mid\sqrt{A_{x}^{2}+A_{y}^{2}}\mid$ between an NV center and a $^{11}B$ nuclear spin, which is given by
\begin{align}
A_{x^{'}}=\mid3\sin\alpha\cos\alpha\mid.
\label{Eq5}
\end{align}

In our polarization transfer mechanism, strain is an asset and polarization transfer could be realized by sweeping magnetic field from negative-to-positive or positive-to-negative. There is no microwave driving and we consider an NV center in the $|m_{s}$=$\pm1\rangle$ subspace coupling with a $^{11}B$ nuclear spin, $|m_{s}$=$0\rangle$ is ignored due to large zero-field splitting $D$. The Hamiltonian can be rewritten as
\begin{align}
H_{\pm1}=E_{\perp}\sigma_{x}+\omega(t)\sigma_{z}
+\frac{\bar{Q}}{2}I_{\tilde{z}}+\sigma_{z}\cdot(\sqrt{3}A_{x^{'}}I_{\tilde{x}}+2A_{z}I_{\tilde{z}}),
\end{align}
where $\sigma_{z}$=$(|+1\rangle\langle+1|-|-1\rangle\langle-1|)$ and $\sigma_{x}$=$(|+1\rangle\langle-1|+|-1\rangle\langle+1|)$.  $E_{\perp}$=$\sqrt{E_{x}^{2}+E_{y}^{2}}$ describes the  perpendicular component of the crystal strain. Because of small magnetic field, the $^{11}B$ nuclear spin has two doubly degenerate states as: $|\uparrow\rangle$=$|m_{I}$=$\pm\frac{3}{2}\rangle$ and $|\downarrow\rangle$=$|m_{I}$=$\pm\frac{1}{2}\rangle$. Analogously, in the effective two-level system of $^{11}B$ nuclear spin, we define the operators $I_{\tilde{z}}$=$\frac{1}{2}(|\uparrow\rangle\langle\uparrow|-|\downarrow\rangle\langle\downarrow|)$ and  $I_{\tilde{x}}$=$\frac{1}{2}(|\uparrow\rangle\langle\downarrow|+|\downarrow\rangle\langle\uparrow|)$.

By considering an NV center interacting with a large number of $^{11}B$ nuclear spins in h-BN nanosheets, the Hamiltonian is as follows
\begin{align}
H_{sys}=E_{\perp}\sigma_{x}+\omega(t)\sigma_{z}
+\sum_{j}\frac{\bar{Q}}{2}I_{\tilde{z}_{j}}+\sum_{j}\sigma_{z}\cdot(\sqrt{3}A_{x^{'}_{j}}I_{\tilde{x}_{j}}+2A_{z_{j}}I_{\tilde{z}_{j}}),
\label{Eq2}
\end{align}
in which $A_{x^{'}_{j}}$ and $A_{z_{j}}$ describe the dipole-dipole interaction between the $j$-th $^{11}B$ nuclear spin and the NV center electron spin, respectively. And the eigenenergies and eigenstates of the NV center are given by
\begin{align}
w_{\pm}&=\pm\sqrt{\omega^{2}(t)+E_{\perp}^{2}},\\
|\chi_{+}\rangle&=\cos\frac{\eta}{2}|+1\rangle+\sin\frac{\eta}{2}|-1\rangle,\\
|\chi_{-}\rangle&=\sin\frac{\eta}{2}|+1\rangle-\cos\frac{\eta}{2}|-1\rangle,
 \end{align}
where $\eta$ is the rotation angle which satisfies the relation $\sin\eta$=$\frac{E_{\perp}}{\sqrt{\omega^{2}(t)+E_{\perp}^{2}}}$.
Go to the new basis $|\chi_{\pm}\rangle$, we find that the system Hamiltonian becomes
\begin{align}
H_{1}&=\omega_{e}\sigma_{\tilde{z}}+\frac{\bar{Q}}{2}I_{\tilde{z}}
+2(\sigma_{\tilde{z}}\cos\eta-\sigma_{\tilde{x}}\sin\eta)\cdot (\sqrt{3}a_{x}I_{\tilde{x}}+2a_{z}I_{\tilde{z}}),
\label{eff1}
\end{align}
with $\sigma_{\tilde{z}}$=$\frac{1}{2}(|
\chi_{+}\rangle\langle\chi_{+}|-|\chi_{-}\rangle\langle\chi_{-}|)$, $a_{x}$ and $a_{z}$ denotes the transverse and parallel effective coupling strength between the NV spin and an ensemble of $^{11}B$ nuclear spins. $\omega_{e}$=2$\sqrt{\omega^{2}(t)+E_{\perp}^{2}}$ is the effective frequency of the NV center.

\begin{figure}[h]
\center\includegraphics[scale=0.6]{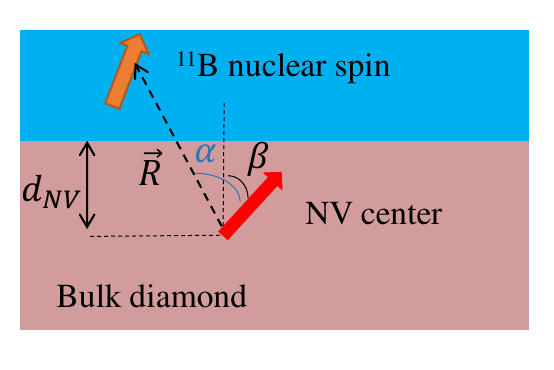}
\caption{An isolated NV center (red arrow) with a depth $d_{NV}$ below the diamond surface interacts with a $^{11}B$ nuclear spin (orange arrow) which is near the surface of diamond. The $\vec{R}$ connects the NV center and the $^{11}B$ nuclear spin, the angle $\alpha$ denotes the angular orientation of $\vec{R}$, and the angle $\beta$ describes quantization axis of the NV center.}
\label{protocol1}
\end{figure}

The magnetic dipole-dipole component $a_{x}$ between an isolated NV center and an ensemble of $^{11}B$ nuclear spins can be estimated as \cite{tetienne2021prospects}
\begin{align}
(a_{x})^{2}=\rho_{n}\int (\frac{\mu_{0}\hbar\gamma_{e}\gamma_{n}}{4\pi R^{3}}A_{x^{'}})^{2}d^{3}\vec{R},
\end{align}
where $A_{x^{'}}$ is effective transverse coupling evaluated analytically in Eq. (\ref{Eq5}). We consider the case of h-BN nanosheets placed on a flat diamond surface interacting with surface shallow NV center shown in Fig. $\ref{protocol1}$, then the effective coupling strength between an NV center and an ensemble of  $^{11}B$ nuclear spins can be written as
\begin{align}
(a_{x})^{2}\approx\frac{4(\mu_{0}\hbar\gamma_{e}\gamma_{n})^{2}\rho_{n}\pi[55+12\cos(2\beta)-3\cos(4\beta)]}{(4\pi)^{2}1024d^{3}_{NV}},
\label{Eq1}
\end{align}
where $^{11}B$ spins in nanosheets offer $80\%$ natural abundance with $\rho_{n}$$=$44 nm$^{-3}$, $d_{NV}$ describes the depth of NV center below the diamond surface (assumed the quantization axis of the NV center $\beta$=$54.7^{\circ}$). Thus, the effective coupling strengths between an NV center and a $^{11}B$ nuclear spin ensemble are approximately $a_{x}$$\approx$$(2\pi)$$\times$11.98 kHz for $d_{NV}$=$5$ nm, $a_{x}$$\approx$$(2\pi)$$\times$25.78 kHz for $d_{NV}$=$3$ nm and $a_{x}$$\approx$$(2\pi)$$\times$47.46 kHz for $d_{NV}$=$2$ nm.

Notice that our polarization transfer mechanism involves NV subspace $|m_{s}$=$\pm1\rangle$ without microwave (MW) driving and doesn't work for spin-$1/2$ nuclear spin species. There are two main reasons. Firstly, at low magnetic field, small Larmor frequencies of spin-$1/2$ nuclear spin species can be comparable or smaller than effective coupling strengthes between the NV spin and nuclei. Additionally, there is no similar hierarchical gaps of NV and $^{13}$C spins encountered in Ref. \cite{ajoy2018orientation}.

\section{Transition probabilities based on Landau-Zener theory}
According Eq. (\ref{eff1}), we have eigenstates $|\chi_{+},\uparrow\rangle$, $|\chi_{+},\downarrow\rangle$, $|\chi_{+},\downarrow\rangle$ and $|\chi_{-},\downarrow\rangle$, and Eq. (\ref{eff1}) can be rewritten as
\begin{eqnarray}
H_{1}&\approx&
 \left(
\begin{array}{cccc}
\omega_{e}+\frac{\bar{Q}}{2} & 0& 0 &0\\
0& \omega_{e}-\frac{\bar{Q}}{2} & \sqrt{3}a_{x}\sin\eta/2 &0\\
  0 & \sqrt{3}a_{x}\sin\eta/2 & -\omega_{e}+\frac{\bar{Q}}{2} &0 \\
0&0&0&  -\omega_{e}-\frac{\bar{Q}}{2}
\end{array}
\right).
\end{eqnarray}
Our scheme starts with a magnetic field initialized to an orientation opposite to that of the NV such that $\omega(t=0)=-(2\pi)\times6$ MHz. Suppose that the NV center electron spin is initially prepared in state $|\chi_{-}\rangle$ (approximated as $|+1\rangle$) through a corresponding microwave (MW) pulse and $^{11}B$ nuclei are unpolarized, i.e., in thermal equilibrium states. If $\bar{Q}\gg a_x, a_z$, the magnetic field is swept to pass through avoided crossings of the energy levels, where the Hartmann-Hahn resonance condition is matched,
\begin{equation}
\omega_{e}=2\sqrt{\omega^{2}(t)+E_{\perp}^{2}}= \frac{\bar{Q}}{2}.
\end{equation}
Therefore, when $E_{\perp}$ is comparable to but not larger than $\frac{\bar{Q}}{4}$, it is possible to have state transition between quantum states $|\chi_{-},\uparrow\rangle$ and $|\chi_{+},\downarrow\rangle$ during the LZ double passage processes, when the state $|\chi_{-},\downarrow\rangle$ is not affected as their dynamics remain adiabatic ($4 E_{\perp}^2/v\gg 1$). Then polarization could be transferred from the NV spin to  $^{11}B$ nuclear spins. The Landau-Zener (LZ) theory describes the transition between two quantum states $|\chi_{-},\uparrow\rangle$ and $|\chi_{+},\downarrow\rangle$ at each of avoided crossing of the energy levels, which is given by in terms of the Landau-Zener formula
\begin{align}
p_{l}&=\exp(-2\pi\mu),
\end{align}
where $\mu$=$\frac{(\frac{\sqrt{3}a_{x}\sin\eta}{2})^{2}}{|2 v\cos\eta|}$ is the adiabatic parameter. The unitary non-adiabatic transformation for these two states is approximately given by the time-independent unitary matrix
\begin{equation}
 U=
 \left(
\begin{array}{cc}
  \sqrt{1-p_{l}}e^{-i\tilde{\varphi}_{s}} & -\sqrt{p_{l}} \\
  \sqrt{p_{l}} & \sqrt{1-p_{l}}e^{i\tilde{\varphi}_{s}}
\end{array}
\right),\\
\end{equation}
where $\tilde{\varphi}_{s}$=$-\frac{\pi}{4}+\mu(\ln\mu-1)+\arg\Gamma(1-i\mu)$, and $\Gamma$ is the Gamma function  \cite{ashhab2007two,shevchenko2010landau}.

According to Hartmann-Hahn resonance condition, when magnetic field is swept from negative-to-positive or positive-to-negative, if we have $0<|E_{\perp}|<\frac{\bar{Q}}{4}$, two sequential avoided crossings induce the probability of the state transfer from $|\chi_{-},\uparrow\rangle$ to state $|\chi_{+},\downarrow\rangle$ as $P_{l}=P\sin^{2}[\Phi]$ with
\begin{align}
P=4p_{l}(1-p_{l}),
\end{align}
where the phase $\Phi$ is acquired during the adiabatic evolution and non-adiabatic transitions \cite{ashhab2007two,shevchenko2010landau}.

\begin{figure}[h]
\center\includegraphics[scale=0.8]{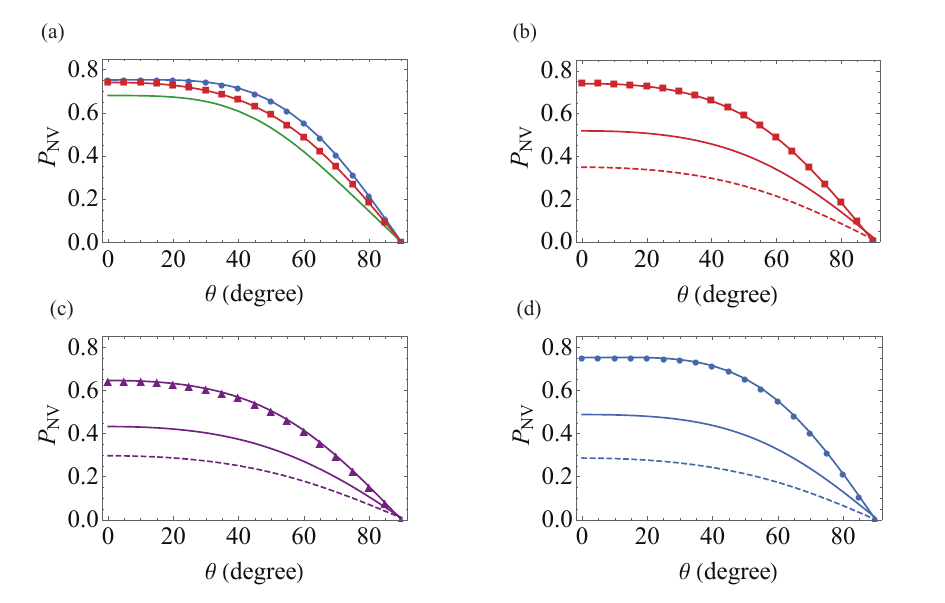}
\caption{(a) The ensemble-averaged initial polarization of the NV centers as a function of the relative angle $\theta$ between their orientations and the applied magnetic field by assuming 300 times averaging in random values within $E_{z}\in(2\pi)\times$[-0.73, 0.73] MHz, the MW Rabi frequency is given by $\Omega=$$(2\pi)\times$ 4(green solid), 6(blue circles), 8(red cubes) MHz. (b) The MW Rabi frequency is given by $\Omega=$$(2\pi)\times$ 8 MHz within $E_{z}\in(2\pi)\times$[-0.73, 0.73] MHz and $\delta_b=(2\pi)\times$0(red cubes), $\in(2\pi)\times$[-2, 2](red solid), [-3, 3](red dashed) MHz. (c) The MW Rabi frequency is given by $\Omega=$$(2\pi)\times$ 8 MHz within $E_{z}\in(2\pi)\times$[-1.5, 1.5] MHz and $\delta_b=(2\pi)\times$0(purple triangles), $\in(2\pi)\times$[-2, 2](purple solid), [-3, 3](purple dashed) MHz. (d) The MW Rabi frequency is given by $\Omega=$$(2\pi)\times$ 6 MHz within $E_{z}\in(2\pi)\times$[-0.73, 0.73] MHz and $\delta_b=(2\pi)\times$0(blue circles), $\in(2\pi)\times$[-2, 2](blue solid), [-3, 3](blue dashed) MHz.}
\label{protocol2}
\end{figure}

\section{Polarization initialization of the nitrogen-vacancy spin in nanodiamonds}
In order to investigate effects of the orientation of the applied magnetic field and demonstrate the robustness to orientations, we define the polarization efficiency of NV center as a function of the relative angle $\theta$ between the applied magnetic field and the NV center axis in nanodiamond powders. The effective Hamiltonian of the NV ground state triplet when a magnetic field $B$ is applied, is given by
\begin{eqnarray}
H_{NV}^{a}&=& (D+E_z)S^{2}_{z}+\gamma_e\boldsymbol{B}\cdot \boldsymbol{S}+\delta_bS_z+ E_{\perp}(S_{x}^{2}-S_{y}^{2}) \notag \\
&=&
 \left(
\begin{array}{ccc}
D+E_z+\gamma_e B_z+\delta_b& \frac{1}{\sqrt{2}}\gamma_e (B_x-i B_y)&E_{\perp}\\
  \frac{1}{\sqrt{2}}\gamma_e (B_x+i B_y)& 0 &  \frac{1}{\sqrt{2}}\gamma_e (B_x-i B_y)\\
 E_{\perp}&  \frac{1}{\sqrt{2}}\gamma_e (B_x+i B_y) & D+E_z-\gamma_e B_z-\delta_b \\
\end{array}\label{Eq27}
\right),
\end{eqnarray}
%
where $\gamma_e$ is the gyro-magnetic ratio of the NV center, $\delta_b$ is  inhomogeneous broadening induced by interactions with electron and nuclear spins nearby the NV center, $B_x$, $B_y$, $B_z$ are the components of the applied magnetic field in different directions. By considering the magnetic field is applied ($\gamma_e B_z\gg E_{x}, E_{z}$) and a large zero field splitting $(D\gg \gamma_e B_x,  \gamma_e B_y,  \gamma_e B_z)$ leads transverse magnetic field parts to be negligible, the NV spin Hamiltonian is approximated to be
\begin{equation}
 H_{NV}^{b}\approx (D+E_z)S^{2}_{z}+(\omega\cos\theta+\delta_b) S_{z},
\end{equation}
with $\omega=\gamma_e B$ and $\theta$ is the angle between the magnetic field orientation and the NV orientation. Similarly, we can have the Hamiltonian of an NV center of a random orientation driven by an MW field to be approximated as
\begin{align}
H_{NV}^{c}\approx (D+E_z)S^{2}_{z}+(\omega\cos\theta+\delta_b) S_{z}+\sqrt{2}\Omega\sin\phi \cos\omega_m tS_{x},
\end{align}
where $\phi$ is the angle between the NV orientation and the microwave field. $\Omega$ and $\omega_m$ are the Rabi and driving frequency of microwave field, respectively. In the interaction picture with respect to $H^{'}_{0}$=$(D+\omega) S^{2}_{z}$, after the rotating wave approximation with the resonance condition $\omega_m=D+\omega$, the Hamiltonian can be obtained as
\begin{equation}
 H_{NV}^{'}=
 \left(
\begin{array}{ccc}
\omega\cos\theta+\delta_b-\omega+E_z & \frac{1}{2}\Omega\sin\phi & 0\\
  \frac{1}{2}\Omega\sin\phi & 0 & \frac{1}{2}\Omega\sin\phi\\
  0 & \frac{1}{2}\Omega\sin\phi & -\omega\cos\theta-\delta_b-\omega+E_z \\
\end{array}\label{Eq27}
\right).
\end{equation}
According to Fermi's Golden rules, the NV spin transition probability from the $|m_{s}$=$0\rangle$ state to the $|m_{s}$=$+1\rangle$ state and from the $|m_{s}$=$0\rangle$ state to the $|m_{s}$=$-1\rangle$ state are as follows
\begin{align}
P_{0\rightarrow +1}=\frac{(\Omega\sin\phi)^{2}}{(\Omega\sin\phi)^{2}+(\omega\cos\theta-\omega+\delta_b+E_z)^{2}}\sin^{2}\frac{\sqrt{(\Omega\sin\phi)^{2}+(\omega\cos\theta-\omega+\delta_b+E_z)^{2}}t}{2},\\
P_{0\rightarrow -1}=\frac{(\Omega\sin\phi)^{2}}{(\Omega\sin\phi)^{2}+(-\omega\cos\theta-\omega-\delta_b+E_z)^{2}}\sin^{2}\frac{\sqrt{(\Omega\sin\phi)^{2}+(-\omega\cos\theta-\omega-\delta_b+E_z)^{2}}t}{2}.
\end{align}
Due to different angles $\phi$ of the microwave pulse, we calculate the average transition probability which is an average over all the possible angles, then the average transition probability from $|m_{s}$=$0\rangle$ $\rightarrow$ $|m_{s}$=$+1\rangle$ and $|m_{s}$=$0\rangle$ $\rightarrow$ $|m_{s}$=$-1\rangle$ are given by
\begin{align}
\overline{P}_{0\rightarrow +1}&=\frac{1}{2}\int^{\pi}_{0}\frac{(\Omega\sin\phi)^{2}}{(\Omega\sin\phi)^{2}+\Delta^{2}}\sin^{2}(\frac{\sqrt{(\Omega\sin\phi)^{2}+\Delta^{2}}t}{2})\sin\phi d\phi,\\
\overline{P}_{0\rightarrow -1}&=\frac{1}{2}\int^{\pi}_{0}\frac{(\Omega\sin\phi)^{2}}{(\Omega\sin\phi)^{2}+L^{2}}\sin^{2}(\frac{\sqrt{(\Omega\sin\phi)^{2}+L^{2}}t}{2})\sin\phi d\phi,
\end{align}
here $\Delta$=$\omega\cos\theta-\omega+E_z+\delta_b$ and $L$=$-\omega(1+\cos\theta)+E_z-\delta_b$. We take $t$=$\pi/\Omega$ and the ensemble-averaged initial polarization of the NV center $P_{NV}$ is given by
\begin{align}
P_{NV}(\theta)=P_i(\overline{P}_{0\rightarrow +1}-\overline{P}_{0\rightarrow -1}),
\end{align}
where we assume that the NV center is fully initialized to $m_s=0$ state and its initial population $P_i=1$. If a magnetic field is initialized to along the NV orientation such that $\omega(t=0)=-(2\pi)\times6$ MHz, the NV center electron spin is initially prepared in state $|\chi_{-}\rangle$ (approximated as $|+1\rangle$), we have the ensemble-averaged initial polarization of NV center $P_{NV}(\theta)$, as shown in Fig. \ref{protocol2}, when MW Rabi frequency is given by $\Omega=$$(2\pi)\times$4(blue dashed), 6(red circles), 8(green dotted) MHz.

\section{Robustness}
We investigate the effects of the coherence time of NV center, the variances of coupling strengths $a_{x}$ and the crystal strains $E_{\perp}$, and the other electron and nuclear spins in the system to polarization transfer between an nitrogen vacancy (NV) center and an effective $^{11}B$ nuclear spin in this section.
\subsection{The influence of the NV coherence time $T_{2}$}
Our detailed numerical simulations accounting for the coherence time $T_{2}(NV)$ of the NV center are shown in Fig. 1(d) in main text, it is reasonable to describe the system by using a Lindblad master equation with the form
\begin{eqnarray}
\dot{\rho}=-i[H_{1},\rho]+L_{e}\rho L^{\dagger}_{e}
-\frac{1}{2}(L^{\dagger}_{e}L_{e}\rho+\rho L^{\dagger}_{e}L_{e}),
\end{eqnarray}
where $\rho$ represents the density matrix of whole system, $H_{1}$ is shown in Eq. (\ref{eff1}), and $L_{e}=\sqrt{\Gamma_{e}}(|+1\rangle\langle+1|-|-1\rangle\langle-1|)$ with $\Gamma_{e}=1/T_{2}(NV)$. In order to investigate the production of the polarization, the polarization is modeled by writing the total density matrix $\rho$ as $\rho_{NV}\otimes Tr_{NV}\{\rho\}$, in which $Tr_{NV}\{\rho\}$ contributes to the partial trace over the NV degrees of freedom. In all simulations, we assume that  $T_{2}(NV)$ is longer than the characteristic LZ time ($\tau$) to coherent transfer and smaller than the polarization magnetic field sweeping period ($T_{w}$), i.e., $\tau<T_{2}(NV)<T_{w}$.

\begin{figure*}[htp]
\center\includegraphics[scale=0.75]{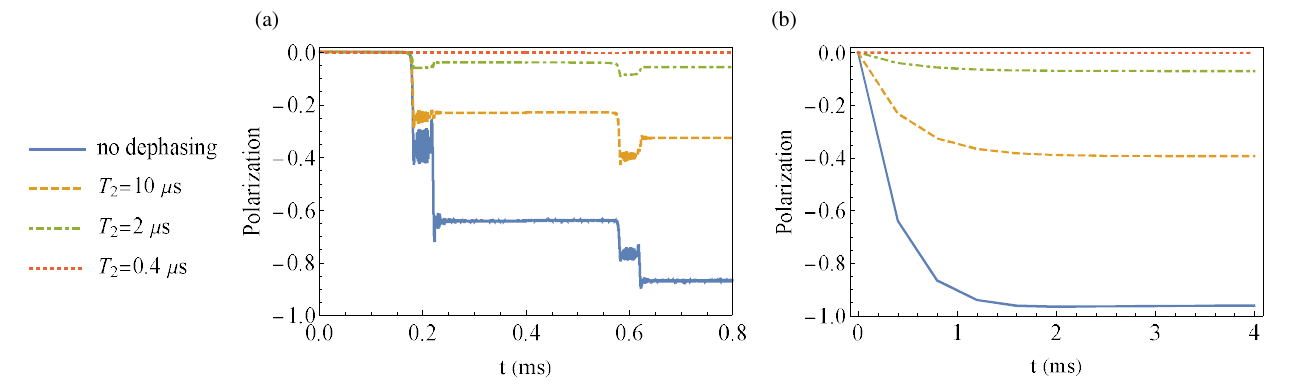}
\caption{(a) Polarizations of $^{11}B$ nuclear spins are shown by considering different coherence time $T_{2}$ of the NV spin. The strength of electric strain $E_{\perp}$$=$$(2\pi)\times$0.4 MHz, the perpendicular coupling between NV and nuclear spins $a_{x}=(2\pi)\times0.05$ MHz and the sweep rate $v$=$(2\pi)$$\times$30 MHz/ms. The full cycle containing negative-to-positive and positive-to-negative sweepings is presented and the initial value of the magnetic field is $w(t=0)=-(2\pi)\times6$ MHz. (b) The total $^{11}B$ polarization about multiple repetitions of the optical excitation and field sweep cycles is described. All of parameters are the same as (a).}
\label{protocol3}
\end{figure*}

To demonstrate the robustness to the $T_{2}(NV)$ in our scheme, we evaluate the polarization between an NV center and $^{11}B$ nuclear spins. Results for different $T_{2}$ are given by orange dashed ($T_{2}=10$ $\mu$s), green dot-dashed ($T_{2}=2$ $\mu$s) and red dotted line ($T_{2}=0.4$ $\mu$s) respectively. We can attain considerable dynamical nuclear polarization efficiency when $\tau<T_{2}(NV)<T_{w}$, see Fig. \ref{protocol3}(a). The whole evolution of the $^{11}B$ polarization as a function of time upon multiple cycles is described in Fig. \ref{protocol3}(b).
%
\subsection{The influence of variance of crystal strains and coupling strengths}
\begin{figure}[h]
\center\includegraphics[scale=0.56]{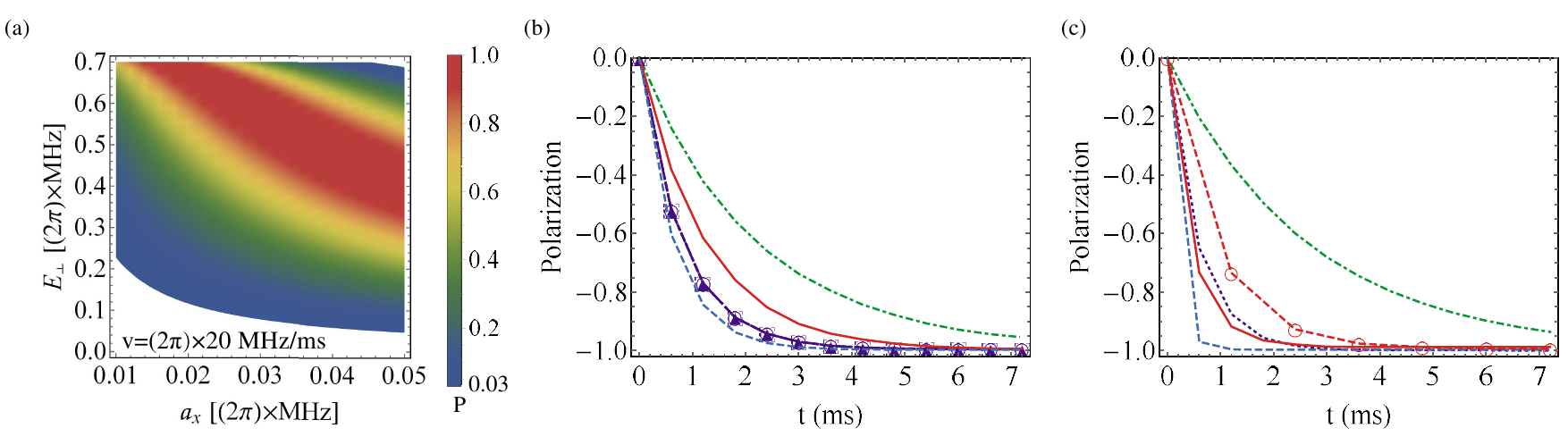}
\caption{Robustness of our mechanism to crystal strains $E_{\perp}$ and coupling strengths $a_{x}$. (a) Transfer probability P as a function of the coupling strengths and crystal strains to show the robustness with the sweep $v$=$(2\pi)$$\times$20 MHz/ms. (b) Polarization is transferred to a $^{11}$B spin by assuming the initial value of the magnetic field $w(t=0)=-(2\pi)\times6$ MHz, the sweep rate $v=$$(2\pi)\times$20 MHz$/$ms, the coupling strength $a_{x}$=$(2\pi)\times$0.04 MHz, and the crystal strains are given as $E_{\perp}$=$(2\pi)\times$0.65(red solid); 0.5(blue dashed); 0.4(purple triangles); 0.26(green dot-dashed) MHz. The purple dashed lines present $E_{\perp}=E_{x}=(2\pi)\times$0.4(hollow circles) MHz and $E_{\perp}=E_{y}=(2\pi)\times$0.4(hollow cubes) MHz. (c) Polarization is transferred to a $^{11}$B spin by assuming $E_{\perp}$=$(2\pi)\times$0.52 MHz, $w(t=0)=-(2\pi)\times6$ MHz, and the coupling strengths are given as $a_{x}$=$(2\pi)\times$0.047(red solid); 0.04(blue dashed);  0.03(purple dotted); 0.02(green dot-dashed) MHz;  $w(t=0)=-(2\pi)\times12$ MHz, $w(t_f)=(2\pi)\times12$ MHz and $a_{x}$=$(2\pi)\times$0.047 MHz (red circles dotted).}
\label{protocol4}
\end{figure}

Here, we investigate the robustness of our scheme to the crystal strains inhomogeneity and variance of coupling strengths, with the sweep rate $v$=$(2\pi)$$\times$20 MHz/ms as shown in Fig. \ref{protocol4}(a). Polarization buildup for different strains $E_{\perp}$ with the same $a_{x}=(2\pi)\times0.04$ MHz are shown in Fig. \ref{protocol4}(b). Our method is robust to the variance of transverse crystal strains, which is not related to crystal strain direction inhomogeneity. For example, as shown in Fig. \ref{protocol4}(b), when the strain satisfies $E_{x}=(2\pi)\times0.4$ MHz and $E_{y}=0$ MHz ($E_{\perp}=E_{x}$) (hollow circles), it owns almost the same performance to the case when $E_{x}=0$ MHz and $E_{y}=(2\pi)\times0.4$ MHz ($E_{\perp}=E_{y}$) (hollow cubes). Our transfer mechanism is also robust to variance of different coupling strengths $a_{x}$ between the NV center and $^{11}B$ nuclear spins, see Fig. \ref{protocol4}(c). Additionally, the system is not sensitive to the exact start and end magnetic field values, see Fig. \ref{protocol4}(c), another start $w(t=0)=-(2\pi)\times12$ MHz also shows the polarization transfer but a larger sweep range leads to a slower polarization transfer.

\subsection{The influence of other spins}\label{Eq6}
In addition to the target nuclear spins $^{11}B$, there are other spins in the system, i.e., the host $^{14}N$ spin of the NV center, the P1 centers in diamonds and $^{14}N$ nuclear spins in h-BN nanosheets. Therefore, it is necessary to take into account the effects of other spins, and the total Hamiltonian is given by
\begin{align}
H_{oth}&=H_{^{14}N}+H_{P1}+H^{'}_{^{14}N}+H_{dip(NV,^{14}N)}+H_{dip(NV,P1)}+H_{dip(^{11}B,P1)}+H^{'}_{dip(NV,^{14}N)}.
\end{align}
\subsubsection{The host $^{14}N$ nuclear spin}
\begin{figure}[h]
\center\includegraphics[scale=0.5]{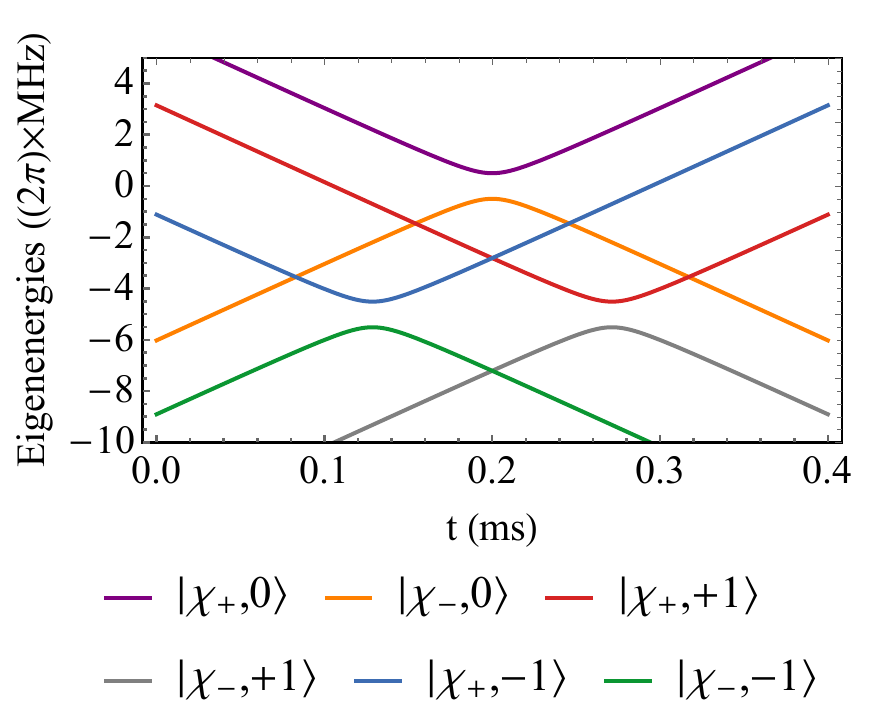}
\caption{The energy levels of the NV center coupling with the host $^{14}N$ nuclear spin based on Eq. (28).}
\label{protocol5}
\end{figure}

The NV-$^{14}N$ interaction Hamiltonian is given by
\begin{align}
H_{dip(NV,^{14}N)}&=\vec{S}\cdot\mathcal{A^{'}}\cdot\vec{I^{'}}\\\nonumber
&\approx S_{z}A_{\parallel}I^{'}_{z}+A_{\perp}(S_{x}I^{'}_{x}+S_{y}I^{'}_{y}),
\end{align}
where parallel and perpendicular hyperfine coupling parameters for the host $^{14}N$ nuclear spin and the NV center are denoted as $A_{\parallel}$$=$$-(2\pi)$$\times$2.14 MHz and $A_{\perp}$$=$$-(2\pi)$$\times$2.7 MHz. The quadrupole coupling constant is $\bar{P}$=$-(2\pi)\times$5.01 MHz for the $^{14}N$ nuclear spins. By considering the effective two-level system of the NV spin and the interaction Hamiltonian $H_{dip(NV,^{14}N)_{eff}}\approx A_{\parallel}\sigma_{z}I^{'}_{z}$, we have the system Hamiltonian to be approximated as
\begin{align}
H_{NV-^{14}N}\approx E_{\perp}\sigma_{x}+\omega(t)\sigma_{z}+\bar{P}I^{'2}_{z}+A_{\parallel}\sigma_{z}I^{'}_{z},
\end{align}
in which the eigenenergies of the hyperfine states are given by
\begin{align}
E_{\chi_{+}, 0}&=\sqrt{\omega^{2}(t)+E_{\perp}^{2}},\\
E_{\chi_{-}, 0}&=-\sqrt{\omega^{2}(t)+E_{\perp}^{2}},\\
E_{\chi_{+}, +1}&=\sqrt{(\omega(t)+A_{\parallel})^{2}+E_{\perp}^{2}}+\bar{P},\\
E_{\chi_{-}, +1}&=-\sqrt{(\omega(t)+A_{\parallel})^{2}+E_{\perp}^{2}}+\bar{P},\\
E_{\chi_{+}, -1}&=\sqrt{(\omega(t)-A_{\parallel})^{2}+E_{\perp}^{2}}+\bar{P},\\
E_{\chi_{-}, -1}&=-\sqrt{(\omega(t)-A_{\parallel})^{2}+E_{\perp}^{2}}+\bar{P}.\\\nonumber
\end{align}
As shown in the Fig. $\ref{protocol5}$ here and Fig. 1(f) in main text, the magnetic field is swept to pass different energy level anti-crossings, and the hyperfine interaction of the NV center electron spin with the host $^{14}N$ nuclear spin does not affect polarization transfer efficiency.

\subsubsection{The P1 centers}
\begin{figure}[h]
\center\includegraphics[scale=0.45]{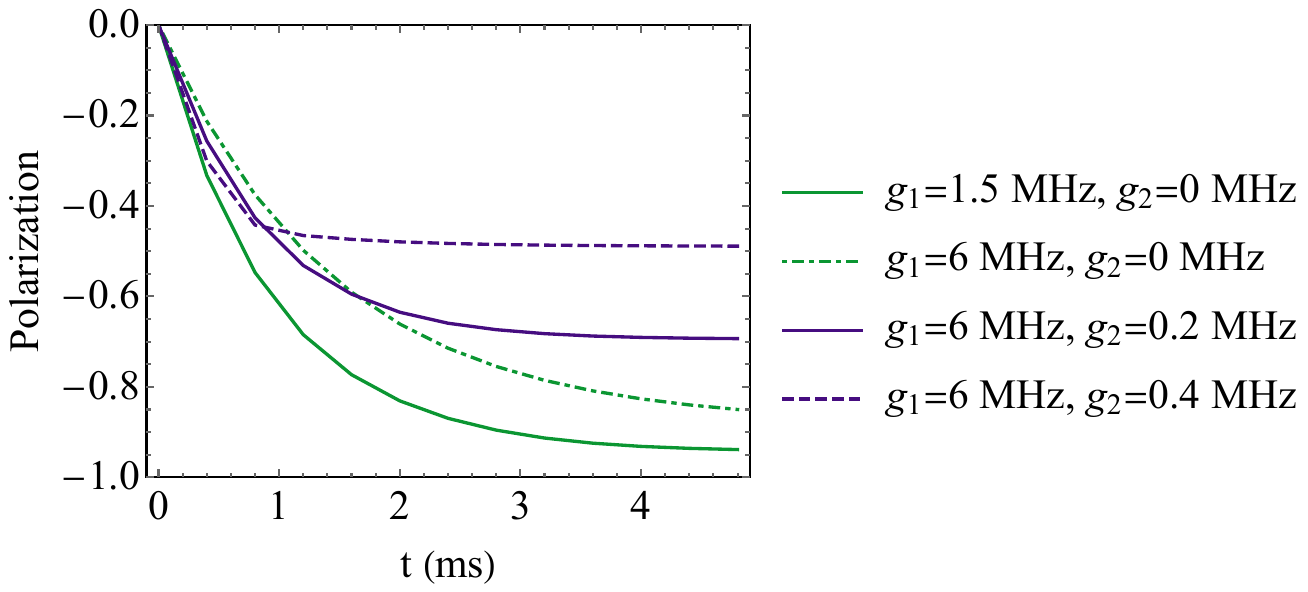}
\caption{ Polarizations of $^{11}B$ nuclear spins considering the coupling of P1 centers with the NV center and $^{11}$B nuclear spins. By assuming the initial value of the magnetic field $w(t=0)=-(2\pi)\times6$ MHz, the sweep rate $v=$$(2\pi)\times$30 MHz$/$ms, the crystal strains $E_{\perp}$=$(2\pi)\times$0.4 MHz and the coupling strength between the NV center and $^{11}B$ spins $a_{x}$=$(2\pi)\times$0.04 MHz.}
\label{protocol6}
\end{figure}

We investigate the impact of other paramagnetic impurities, i.e., P1 centers (a spin-1/2 defect) on the diamond surface to our scheme. The interaction between P1 centers and the NV center $H_{dip(NV,P1)}$ and the interaction between P1 centers and the $^{11}B$ nuclear spin $H_{dip(^{11}B,P1)}$ are given by the Hamiltonian

\begin{align}
H_{int}&= H_{dip(NV,P1)}+H_{dip(^{11}B,P1)}\\\nonumber
&= g_{1}[3\frac{(\vec{S}\cdot\vec{R}_{1})(\vec{S^{'}}\cdot\vec{R}_{1})}{R^{2}_{1}}-\vec{S}\cdot\vec{S^{'}}]+g_{2}[3\frac{(\vec{S^{'}}\cdot\vec{R}_{2})(\vec{I}\cdot\vec{R}_{2})}{R^{2}_{2}}-\vec{S^{'}}\cdot\vec{I}],
\end{align}
where the spin operators $\vec{S}$, $\vec{S^{'}}$ and $\vec{I}$ correspond to the NV center electron spin, the P1 center electron spin and the $^{11}B$ nuclear spin, respectively. $R_{1}=|\vec{R}_{1}|$ stands for the distance between the NV center electron spin and the P1 center electron spin. Analogously, $R_{2}=|\vec{R}_{2}|$ stands for the distance between the $^{11}B$ nuclear spin and the P1 center electron spin. $\alpha_{1,2}=45^{\circ}$ and $\varphi_{1,2}=45^{\circ}$ define the angular orientation of $\vec{R}_{1,2}$.  $g_{1}$ is the dipolar coupling constant of the NV center electron spin and the P1 center electron spin, and $g_{2}$ is the dipolar coupling constant of the $^{11}B$ nuclear spin and the P1 center electron spin. By assuming that there is a P1 center near the NV center but far away from the $^{11}B$ nuclear spin, the impact of the kind of P1 center is similar to the impact from the host $^{14}N$ spin of the NV center, i.e., it shows polarization build-up when $g_{1}=(2\pi)\times6$ MHz (corresponding to spin distances $\sim$2.05 nm), see Fig. $\ref{protocol6}$. However, polarization transfers might be disturbed when the P1 center is also near the target $^{11}B$ nuclear spins (spin distances $<$1 nm).

\section{The polarization of $^{14}N$ nuclear spins in h-BN nanosheets}
\begin{figure}[h]
\center\includegraphics[scale=0.5]{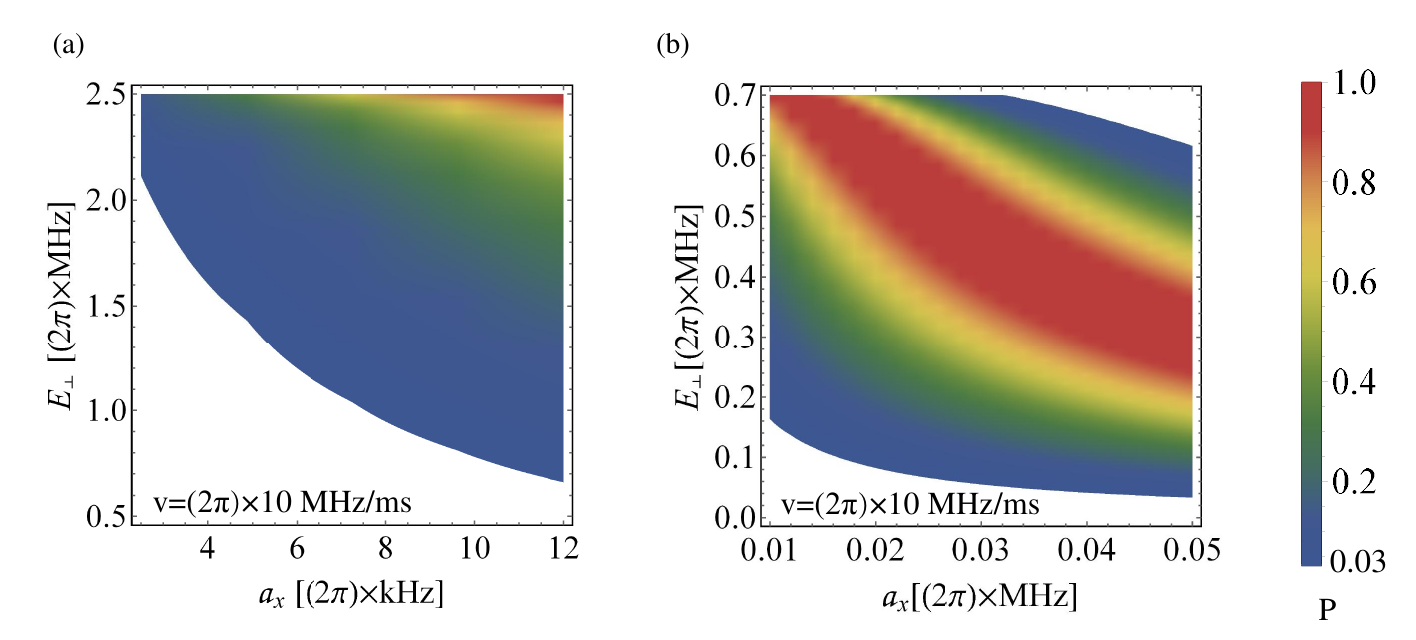}
\caption{ Transfer probability P as a function of the coupling strengths and crystal strains to show the robustness with the sweep rate $v$=$(2\pi)$$\times$10 MHz/ms, when we consider $^{14}N$ spins (a) and $^{11}B$ spins (b) in nanosheets as our polarizing targets.}
\label{protocol7}
\end{figure}
\begin{figure}[htp]
\center\includegraphics[scale=0.62]{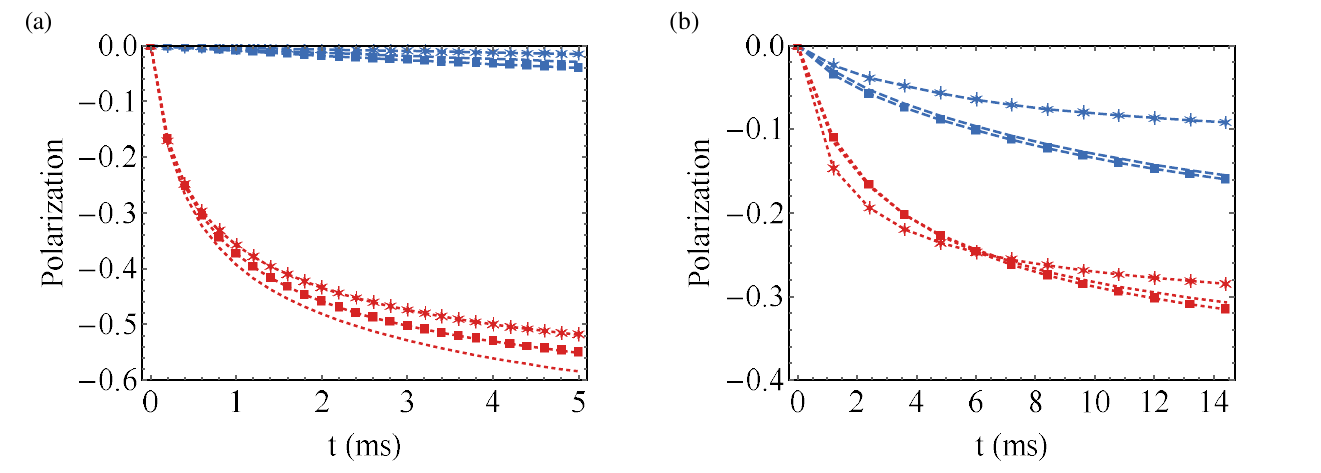}
\caption{ Polarization is transferred to $^{14}N$ spins and $^{11}B$ spins by assuming the initial value of the magnetic field $w(t=0)=-(2\pi)\times6$ MHz, the coupling strength between the NV center and $^{14}N$ spins $a_{x}^{'}$=$(2\pi)\times$10 kHz, and the coupling strength between the NV center and $^{11}B$ spins $a_{x}$=$(2\pi)\times$0.047 MHz. (a) The sweep rate $v=$$(2\pi)\times$60 MHz$/$ms, the crystal strains $E_{\perp}$$\in$$(2\pi)\times$[-1.5, 1.5] MHz, and the random magnetic field noise $\delta_b$$\in$$(2\pi)\times$[-3, 3] MHz. The blue lines present that polarisation is transferred to $^{14}N$ spins with different angles $\theta$ between NV's orientation and magnetic field with $\theta=0^{\circ}$ (dashed line), $40^{\circ}$ (cubes) and $75^{\circ}$ (stars). The red lines present that polarization is transferred to $^{11}B$ spins with different angles $\theta=0^{\circ}$ (dotted line), $40^{\circ}$ (cubes) and $75^{\circ}$ (stars). (b) The sweep rate $v=$$(2\pi)\times$10 MHz$/$ms, the crystal strains $E_{\perp}$$\in$$(2\pi)\times$[-3, 3] MHz, and the random magnetic field noise $\delta_b$$\in$$(2\pi)\times$[-3, 3] MHz. The blue lines present that polarization is transferred to $^{14}N$ spins with different angles $\theta=0^{\circ}$ (dashed line), $40^{\circ}$ (cubes) and $75^{\circ}$ (stars). The red lines present that polarization is transferred to $^{11}B$ spins with $\theta=0^{\circ}$ (dotted line), $40^{\circ}$ (cubes) and $75^{\circ}$ (stars).}
\label{protocol9}
\end{figure}

There are abundant $^{14}N$ nuclear spins in h-BN nanosheets as well and the quadrupole coupling constant of $^{14}N$ nuclear spins is $\bar{P}=-(2\pi)\times$5.01 MHz. It is necessary to discuss whether our mechanism can transfer the polarization of the NV center to the $^{14}N$ nuclear spins in nanosheets. There are two major differences between these two polarizing targets( $^{14}N$ and $^{11}B$ nuclear spins): the gyromagnetic ratio of $^{14}N$ nuclear spins is much smaller than $^{11}B$ spins and quadrupole coupling constant of $^{14}N$ nuclear spins is larger than  $^{11}B$ spins. Take the distance between the NV center and nanosheets $\sim 2$ nm as an example, we investigate the nuclear polarization efficiency as a function of the angle $\theta$ between NV's orientation and magnetic field, the $\delta_b$ is inhomogeneous broadening induced by interactions with electron and nuclear spins nearby the NV center, and the $E_{\perp}$ is induced by a local deformation of the diamond crystal with inhomogeneous broadening. There is almost no polarization build-up of $^{14}N$ nuclear spins when we take the same parameters in main text with the sweep rate  $v=$$(2\pi)\times$60 MHz$/$ms and average 300 runs to have the averaged polarization transfer of $^{14}N$ spins and $^{11}B$ spins by considering $\delta_b\in(2\pi)\times$[-3, 3] MHz and $E_{\perp}\in(2\pi)\times$[-1.5, 1.5] MHz satisfy the Gaussian probability distributions, see Fig. \ref{protocol9}(a). A smaller sweep rate could lead to polarization transfer to $^{14}N$ nuclear spins as well, as shown in Fig. \ref{protocol7} and \ref{protocol9}(b).
%